\documentclass[aps,prl]{revtex4}

\usepackage{graphicx}

\begin{document}


\title{Predictive Dynamical Systems}


\author{T. Ohira} 
\email[]{ohira@csl.sony.co.jp}
\homepage[]{www.csl.sony.co.jp/person/ohira}

\affiliation{Sony Computer Science Laboratories, Inc., Tokyo, Japan 141-0022}


\date{\today}

\begin{abstract}
We propose and study a system whose dynamics are governed by predictions of its future states. 
General formalism and concrete examples are presented. We find that the dynamical characteristics depend on
both how to shape predictions as well as how far ahead in time to make them. Comparisons of these predictive dynamical
models and corresponding delayed dynamical models are discussed. 
\end{abstract}


\maketitle


Predictive behaviors are common in our everyday activities. A few examples include the timing of braking when driving a car, 
catching a ball, trading stock, and shaping population control policies.                                                                                                                                                                                                                                                                                                             
One of the main principles of normal dynamics is that the past and present decide the future. Most physical theories have
been founded on this principle. It should be noted, however, that the picture of identifying a positron going forward in time
with an electron ``coming back from the future" has brought new insight to elementary particle physics\cite{feynman1961}. 
Given the common occurrence of making predictions, considering theoretical pictures of systems whose dynamics are explicitly
governed by predictions or estimations of future states may be constructive.
In light of this background and motivation, the main theme of this letter is to propse predictive dynamical systems
by presenting concrete examples. The behavior of such dynamical systems depends on both how the predictions are made and 
how far in advance they are made. 

We start with the general differential equation form of predictive dynamical systems, given by\begin{equation}
{dx(t) \over dt} = F(x(t),\bar{x}(\bar{t})).
\end{equation}
Here, $x(t)$ is the dynamical variable, and $F(x)$ is the dynamical function. $\bar{x}(\bar{t})$ is a prediction of $x$ at future time $t<\bar{t}$.
This dynamical equation implies that the rate of change of $x(t)$ depends not only on its current state, but also on the predicted future state $\bar{x}(\bar{t})$ 
through dynamical function $F$.
Naturally, when $\bar{t}=t$, it reduces to a normal dynamical equation. In this letter, we discuss the class of equations of the form
\begin{equation}
{dx(t) \over dt} = -\alpha x(t) + f(\bar{x}(\bar{t}))
\label{pde}
\end{equation}
with a constant $\alpha > 0$ for comparison with the corresponding delayed dynamical equations
\cite{mackeyglass1977,cookegrossman1982,glassmackey1988,milton1996}.

There is a variety of ways we can choose $\bar{t}$ and the prediction $\bar{x}(\bar{t})$. In this letter, we set $\bar{t} = t + \eta$ with a 
parameter $\eta$, which we call ``advance." In other words, the dynamics are governed by the predicted state of the dynamical variable
$x$ at a fixed interval $\eta$ in the future. To predict $x$, we consider two cases. The first case is to extrapolate the dynamics for
the duration of the advance $\eta$ with normal ($\eta=0$) dynamics, so that
\begin{equation}
\bar{x}(\bar{t}=t+\eta) = \int_{t}^{t+\eta}{F(x(s))}ds + x(t).
\end{equation}
We call this the ``extrapolate prediction." The second case is to assume the current rate of change of $x$ continues for the duration of the advance.
This case is termed the ``fixed rate prediction" and is given by 
\begin{equation}
\bar{x}(\bar{t}=t+\eta) =\eta {dx(t) \over dt}+ x(t).
\end{equation}

We examine how these different predictions, together with the value of the advance, $\eta$, affect the nature of dynamics.  We start our investigations by computer simulation. 
To avoid ambiguity and for simplicity, time-discretized map 
dynamical models, which incorporate the above--mentioned general properties of the predictive dynamical equations, are studied for the rest of this letter.
The general form of the predictive dynamical map is given as follows.
\begin{equation}
x(t+1) = (1-\mu)x(t) + f(\bar{x}(t+\eta))
\end{equation}
where $\mu$ is a rate constant. The extrapolate prediction can be obtained by iterating the corresponding normal map ($\eta=0$) for the duration of $\eta$.
The fixed rate prediction is obtained by setting $\bar{x}(\bar{t}=t+\eta) = \eta (x(t)-x(t-1)) + x(t)$.

\begin{figure}
  \includegraphics[width=.6\textwidth]{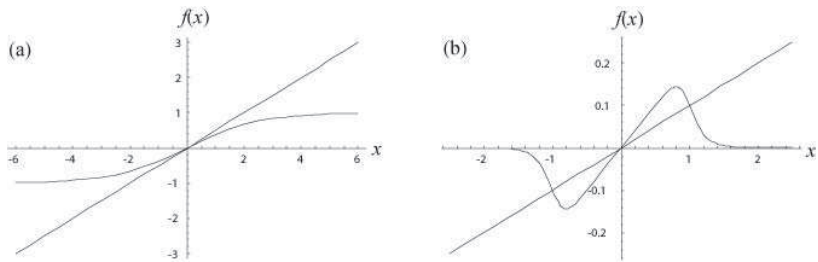}
  \caption{Dynamical functions $f(x)$ with parameters as examples of simulations presented in this letter. (a) Sigmoid function with $\beta = 0.8$. 
Straight line has slope of $\alpha = 0.5$. (b) Mackey-Glass function with parameters $\beta = 0.8$ and $n = 10$. Straight line has 
slope of $\alpha = 0.1$.}
 \end{figure}

The first model we consider is a ``sigmoid" map (Figure 1(a)) with

\begin{equation}
f(x) = {2 \over {1 + e^{-\beta x}}} - 1.
\end{equation}

This function is often used in the context of neural network modelings\cite{cowansharp1989,hertz1991}. We simulate this model with
both the extrapolate and fixed rate predictions. Some examples are shown in Figure 2. In these examples, we have set the 
parameters $\mu$ and $\beta$ to the same values for both prediction schemes. In this parameter set, the origin $x=0$ is a stable fixed point when there is no advance, $\eta=0$.
In the case of the extrapolate prediction, this property is kept even when $\eta$ is increased. 
The situation is quite different for the
case of fixed rate predictions. Here, an increasing $\eta$ breaks the stability of the origin, and periodic behaviors arise.

 \begin{figure}
  \includegraphics[width=.8\textwidth]{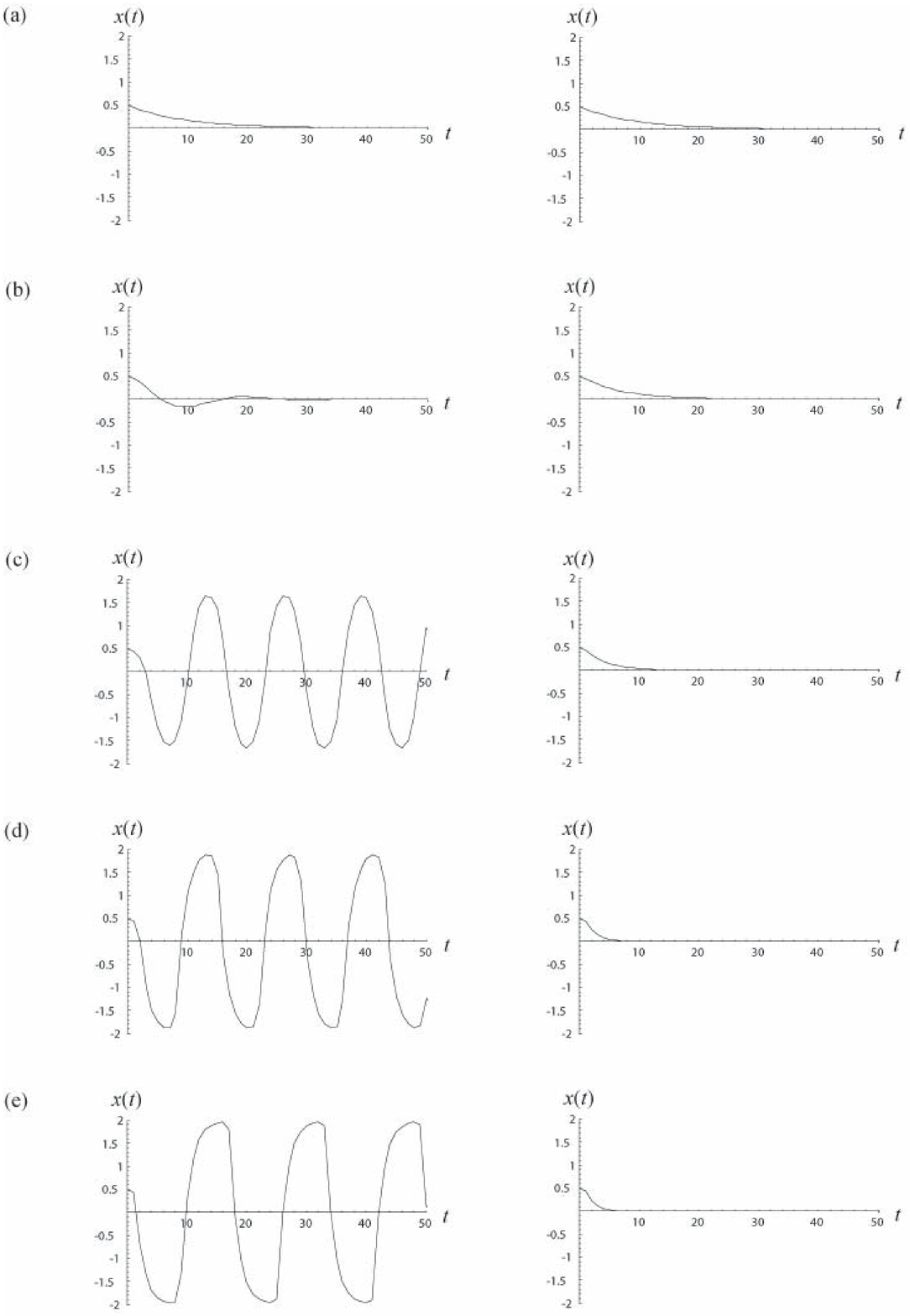}
  \caption{Examples for comparison of fixed rate (left column) and extrapolate (right column) predictions for 
sigmoid map with $\mu=0.5$ and $\beta = 0.8$.
Values of advance $\eta$ are given as (a) 0, (b) 2, (c) 5, (d) 20, and (e) 80.}
 \end{figure}

The same comparison is made with the second model by setting the dynamical function to the ``Mackey-Glass" map (Figure 1(b))  given by
\begin{equation}
f(x) ={ {\beta x} \over {1 + x^{n}}}.
\end{equation}
This function is first proposed in modeling the cell reproduction process and is known to induce chaotic behaviors with a large delay\cite{mackeyglass1977}.
The examples of results from computer simulations are shown in Figure 3. We can see again that even though the extrapolate prediction 
does not change the stability of the fixed point with an increasing advance, the fixed rate prediction case gives rise to complex dynamical behaviors.

 \begin{figure}
  \includegraphics[width=.8\textwidth]{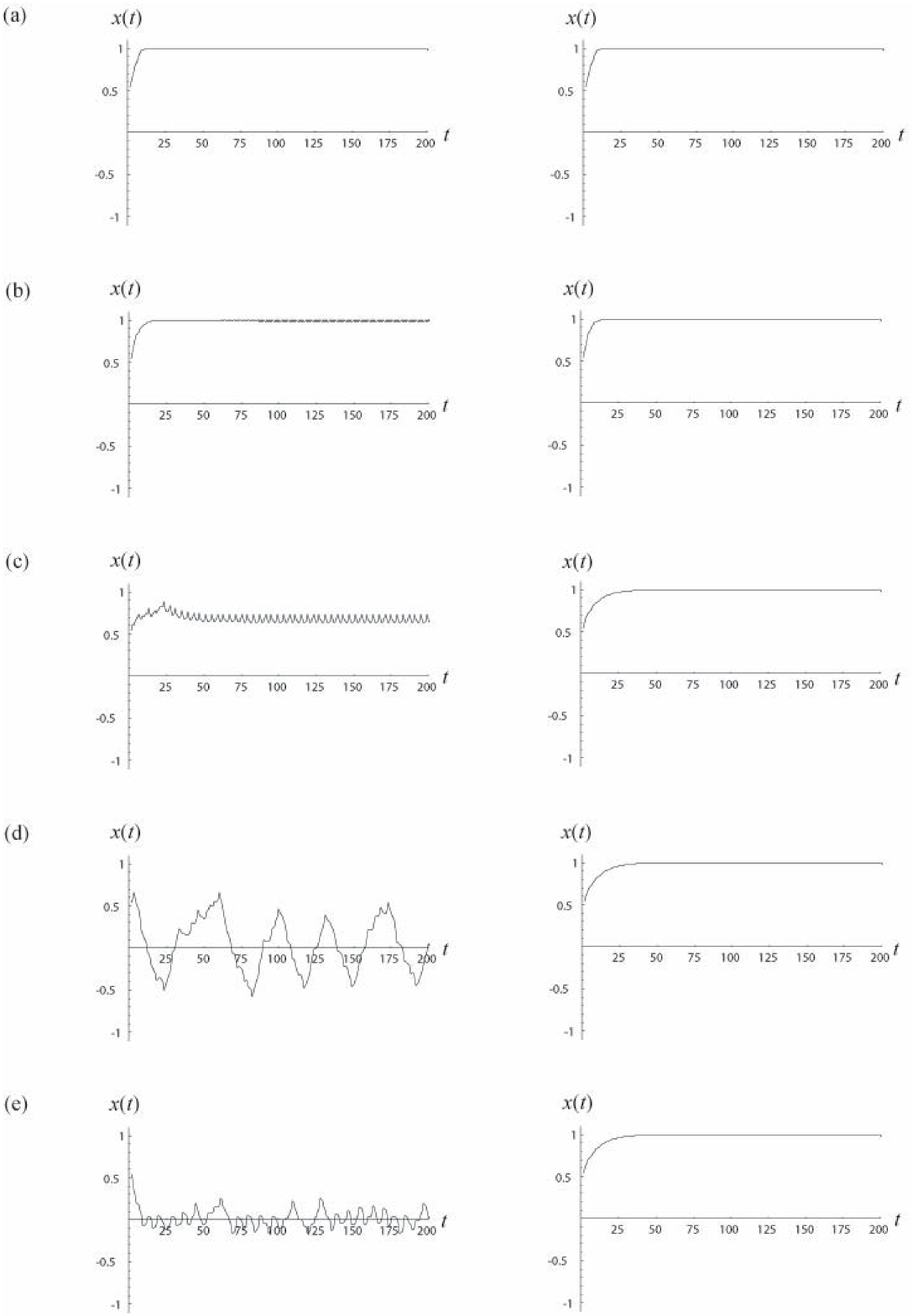}
  \caption{Examples for comparison of fixed rate (left column) and  extrapolate (right column) predictions for 
Mackey-Glass map with $\mu = 0.5$, $\beta = 0.8$, and $n = 10$.
Values of advance $\eta$ are given as (a) 0, (b) 2, (c) 8, (d) 10, and (e) 20.}
 \end{figure}

Now we would like to discuss a couple of issues based on our results on predictive dynamical models.
First, let us examine the differences in dynamical behaviors between the extrapolate and the fixed rate predictions. 
Analytically, we can expand Eq. (\ref{pde}) around the fixed point to examine its stability. 
We can obtain the following through linear stability analysis:
\begin{equation}
(1-\beta\eta){dz(t) \over dt} = (-\alpha+\beta)z(t), \quad z \equiv x - x^{*}, \quad \beta = {df \over dx}{|_{x=x^{*}}}
\end{equation}
where $x^{*}$ is the fixed point. For the case of the fixed rate 
prediction,  we can see that the advance, $\eta$, can switch the stability.  In the extrapolate prediction, on the other hand, 
the stability is not affected by the advance, provided that the corresponding normal dynamics have a monotonic approach to
the stable fixed point. (Details of this stability analysis will be discussed elsewhere\cite{ohira2006}.)
Qualitatively, we can argue that the fixed rate predictions tend to ``overshoot"
compared to the extrapolation, leading to destabilization of the fixed point with a larger advance, $\eta$. 
Higher order analysis and other analytical tools to understand these types of equations need to be developed
to capture the behaviors of dynamics.

Second, we can compare our results with the case of delayed dynamics. Specifically, we
consider the delayed dynamical differential equation and the corresponding map given by
\begin{equation}
{dx(t) \over dt} = -\alpha x(t) + f(x(t-\tau))
\label{dde}
\end{equation}
\begin{equation}
x(t+1) = (1-\mu)x(t) + f(x(t-\tau)).
\end{equation}

These equations describe dynamics whose rate of change is governed by both current and
past states with a fixed interval of delay $\tau$. These equations have been studied with
a variety of applications for systems with delayed feedback.
There are differences and similarities in the predictive and the delayed 
dynamical equations (\ref{pde}) and (\ref{dde}).
We first note that simple replacement of the advance by the delay does not lead to the same characteristics of the dynamics. For example, 
if we simulate both sigmoid and Mackey-Glass maps with the same parameter set as in Figures 2 and 3, and include a delay $\tau = \eta$, the sigmoid case
does not show oscillatory behavior. On the other hand, the Mackey-Glass map shows qualitatively similarly complex behavior with an increasing delay (Figure 4). 
In the case of delayed dynamics, we need to decide on the initial function and delay. Analogously, in predictive dynamics,
the prediction scheme and advance need to be specified. Common to delayed and predictive dynamical systems, both factors respectively affect the nature
of dynamics.

Finally, in the same way that we have considered random walks with delay (delayed random walks)\cite{ohiramilton1995,ohirayamane2000}, random walks with prediction (predictive random walk) can also be considered.
Even with a small delay, delayed random walks give rather complex analytical expressions for statistical quantities, such as variance.  Analogously, the
analysis of predictive random walks is not straightforward\cite{ohira2006}. Mathematically, one may argue that these predictive dynamical and stochastic models can
be cast into the framework of normal non-linear dynamical systems as, after all,  predictions are based on current and past states. Indeed, we can apply
linear stability analysis to gain a partial understanding. 
However, in theoretical modeling, particularly for such fields as physiological controls, economical or social behaviors, and  ecological studies, explicitly taking future predictions into account may be useful. 
explicitly. For example, a recent study of human stick balancing tasks on a fingertip has revealed that corrective motion of the stick is frequently shorter than the human response time\cite{cabreramilton2002,cabreramilton2004a,cabreramilton2004b}. This is a task where both the feedback delay and predictions are intricately
mixed. Models based on delayed dynamics have been proposed, but such models may be further developed by using 
 predictive factors to investigate the experimental results. Models for this and other concrete applications have yet to be constructed 
and further analysis of predictive dynamical systems has yet to be
explored.

  \begin{figure}
  \includegraphics[width=.8\textwidth]{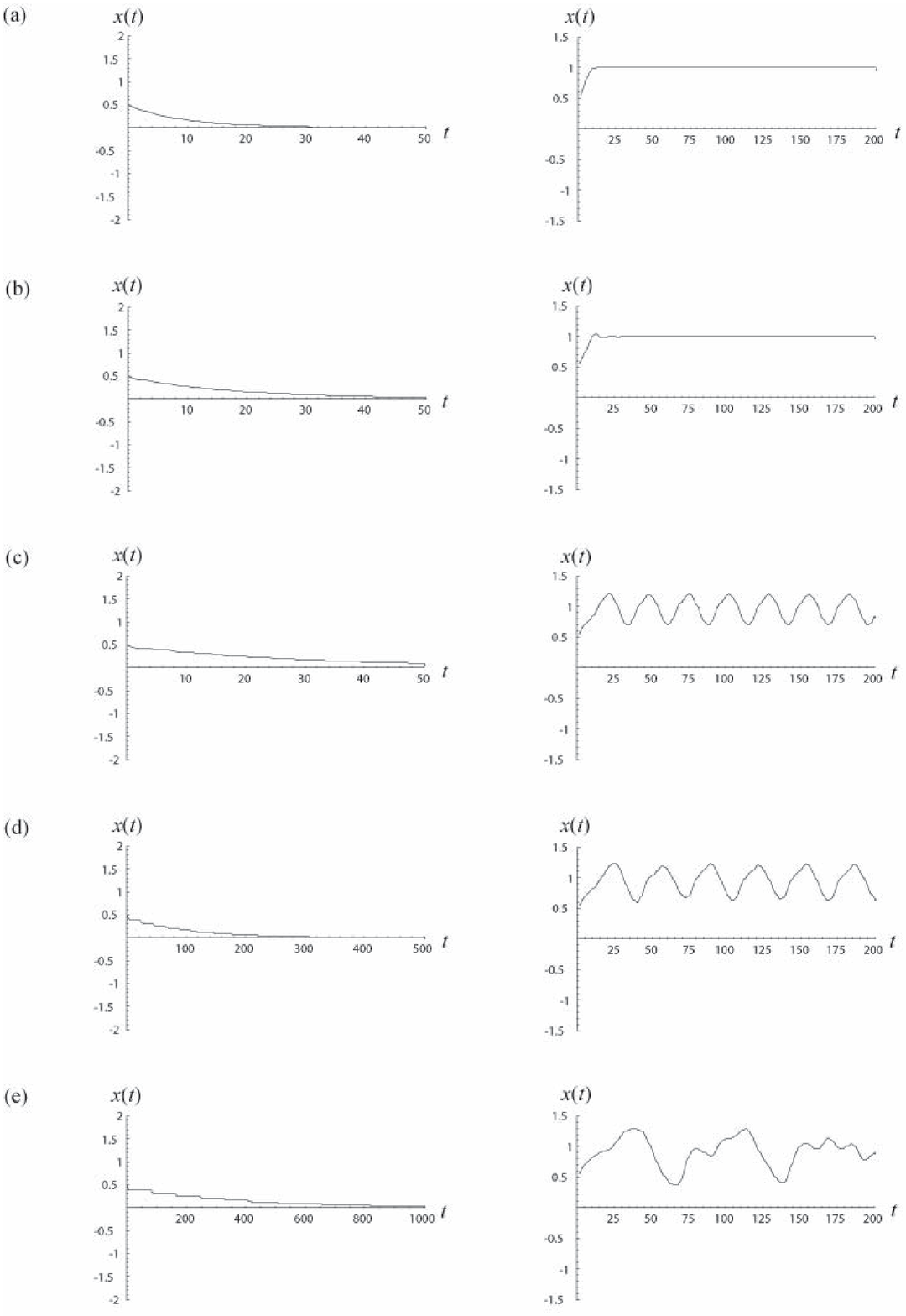}
  \caption{Dynamical behaviors of delayed dynamics for sigmoid (left column) and Mackey--Glass (right column) maps.
Values of delay are set to  $\tau=\eta$ for comparison with previous Figures 2 and 3. For sigmoid map, they are
$\tau=$ (a) 0, (b) 2, (c) 5, (d) 20, and (e) 80, and for Mackey--Glass map, $\tau=$(a) 0, (b) 2, (c) 8, (d) 10, and (e) 20.
(Initial condition during
$(-\tau,0)$ is set at same initial value as $x(0)=0.5$ in predictive cases. ) }
 \end{figure}

\end{document}